\documentclass[12pt]{article}
\usepackage[utf8]{inputenc}
\usepackage{amsmath,amssymb,latexsym}
\usepackage[font=small,labelfont=bf,   justification=justified,   format=plain]{caption}
\usepackage{subcaption}
\usepackage{adjustbox}
\usepackage{enumerate}
\usepackage{graphics}
\usepackage{graphicx}
\usepackage{epsfig}
\usepackage{tikz}
\usepackage[square,sort&compress,comma,numbers]{natbib}
\usepackage{float}
\usepackage{fancyhdr}
\usepackage[a4paper, total={6in, 8in}]{geometry}
\usepackage{booktabs}
\usepackage{makeidx}
\usepackage{amssymb}
\usepackage{eurosym}
\usepackage{placeins}
\usepackage{url}
\usepackage{amsthm}
\usepackage{times}
\usepackage[T1]{fontenc}
\usepackage[utf8]{inputenc}
\usepackage{grffile}
\usepackage{soul} %editing
\graphicspath{{images/}}
\floatstyle{plaintop}
\restylefloat{table}

\newlength\imagewidth
\newlength\imagescale

\emergencystretch=\maxdimen
\begin{document}

% \begin{titlepage}
\title{\bf The Effect of Symmetry Breaking in Coupled Cavity Photonic Crystal Waveguide on Dispersion Characteristics }
\author{Hasan Oguz$^{1,}$ , Zekeriya Mehmet Yuksel$^{1}$, Ozgur Onder Karakilinc$^{2,}$\footnote{okarakilinc@pau.edu.tr},\\ Halil Berberoglu$^{3}$, Mirbek Turduev$^4$, Sevgi Ozdemir Kart$^{1}$, Muzaffer Adak$^{1}$ \\
  {\small $^1$ Department of Physics, Faculty of Science,  }\\
  {\small  Pamukkale University, 20160 Pamukkale, Denizli, Turkey}\\
  {\small $^2$Department of Electric Electronic Engineering, Faculty of Engineering,} \\
  {\small  Pamukkale University, 20160 Pamukkale, Denizli, Turkey} \\
  {\small $^3$ Department of Physics, Polatli Faculty of Science and Letters, } \\
   {\small Ankara Haci Bayram Veli University, 06900 Polatli, Ankara, Turkey }\\
  {\small $^4$ Electrical and Electronics Engineering, Faculty of Engineering, } \\
   {\small Kyrgyz-Turkish Manas University, Bishkek, Chuy, 720038 Kyrgyz Republic }\\
 }

\date{}
\maketitle
 \thispagestyle{empty}

\begin{abstract}
\noindent

In this study, we explore the effect of integrated auxiliary rods at varying angles to the primary cavity rod on the dispersion characteristics of the photonic crystal coupled cavity waveguide (PC CCW). Here, it is intended to break the symmetry of the cavity region by introducing auxiliary rods which gives the degree of freedom for tuning effective index of the PC waveguide. Furthermore, rotational angle variations of auxiliary rods reveal slow light operation of the PC CCW where the group index is maximized and group velocity dispersion, as well as the third-order dispersion, are minimized. In addition, auxiliary rods with a broken symmetry increase not only group index but also operating bandwidth and accordingly increase group bandwidth product by 675\%. Leveraging these results, we demonstrate effective rainbow trapping by manipulating the auxiliary rod angles in photonic crystal coupled cavity waveguides. Our results have encouraging implications for optical buffering, multiplexing, demultiplexing, advanced time-domain and spatial signal processing.

%\bigskip

\noindent
%PACS numbers: 04.50.Kd, 11.15.Kc, 02.40.Yy \\ 

% 11.15.Kc    (Gauge field theory) Classical and semiclassical techniques
% 04.50.Kd    Modified theories of gravity
% 02.40.Yy    Geometric mechanics
\end{abstract}
% \end{titlepage}

\section{Introduction}

Cavity structures hold significant importance in photonics, offering versatile functionalities such as optical filtering, modulation, buffering, delaying and switching within nanophotonic integrated devices and networks  \cite{Yablonovitch1993, Joannopoulos2008, Xia2007, Zhang2008, notomi2008, Xiao2007, Kubo2007, Baba2008, Shu2015}. Coupled cavity waveguide (CCW) photonic crystal (PC) structures, which exploit the slow light (SL) effect with the low group velocity phenomena, emerge as promising avenues for enhancing light-matter interactions in various photonic applications \cite{Kubo2007, Baba2008, Shu2015,notomi2001, Ustun2010}. These CCW PC structures, owing to their compatibility with silicon-on-insulator devices, optical delay lines and electrically integrated components, as well as their compact sizes, offer unique advantages in the realization of photonic integrated circuits \cite{notomi2001}.

The SL phenomenon is one of the core concepts in photonic crystals (PCs) and can be obtained in the flat regions of dispersion curves, which enables the tuning of optical properties \cite{povinelli2005, GIDEN2013}. It is also crucial in realizing optical delay lines, optical buffers, and wavelength demultiplexing devices, among others \cite{Li2008,khatibi2013, Hao2010, Kubo2007, Bagci2015, Danaie2018}. The SL effect can be tuned and enhanced by introducing line defects or manipulating structural defects such as the radius of rods/holes and the shapes of adjacent rows in PC waveguides \cite{Schulz2010,he2018, Moghaddam2019}. Additionally, it is known that low-symmetry PC structures can be exploited to enhance optical properties within the SL region \cite{Shanhui2002, Schulz2010, Yasa2017, Gumus2019, Erim2019, He2021}. Furthermore, the sensitivity of cavity behavior to minor distortions in their geometry presents a powerful tool for optimizing cavity parameters \cite{Vucic2001,Gumus2019}. It is important to note that cavities can also be exploited to demultiplex incoming signals with fine-tuned coupling parameters \cite{Gao2020}.

Another application of SL is the rainbow trapping (RT) effect \cite{Yang2015}. The RT phenomenon involves the temporary trapping of electromagnetic waves with different wavelengths at distinct spatial positions. This effect is primarily attributed to spatial dispersion \cite{Gao2012}. The RT effect can be achieved using graded-index PC structures, tapered PC structures, metallic gratings, chirped and topologically modified PCs, and metamaterials \cite{Tsakmakidis2007,Gan2009, Liu2017, Liu2018, ArreolaLucas2019, Neseli2020, Ghaderian2021}. With the aid of a coupler, point defect, or tapering mechanism, this effect can be implemented in photonic device structures \cite{Hayran:16,Neseli2020}. These structures can be utilized in various applications such as optical switching, multiplexing, demultiplexing, and many other optical communication and information devices \cite{Hayran:16,He2021, Neseli2020,Ghaderian2021}. The adaptability of these structures in photonic devices is pivotal, offering significant enhancements in the efficiency and performance of optical systems.

In this study, to the best of our knowledge, we present the design of rainbow trapping by utilizing the low-symmetry effect in the cavity, rather than in the entire structure, by adding auxiliary rods for the first time. It is important to note that we achieved the RT effect by varying the angles of the auxiliary rods and combining them through symmetry breaking, without tapering or grading the PC structure. Our methods are based on plane wave expansion (PWE) via MPB and finite-difference time-domain (FDTD) via Lumerical.

\section{Design Approach and Numerical Analysis of Coupled Cavity PC Waveguide with Symmetry Breaking}

In this study, we present a detailed exploration of the impacts of symmetry reduction on the dispersion curves of waveguides exhibiting the SL effect. We employ both the plane wave expansion (PWE) and finite-difference time-domain (FDTD) methods for frequency and time domain analysis of the photonic crystal (PC) structures to reveal the SL effect and apply it to the rainbow trapping application, respectively \cite{mpb,taf05,meep}. The combination of these approaches provides a comprehensive exploration of the influence of symmetry reduction on SL properties in both the time and spectral domains. Throughout the study, freely available software MPB (based on the PWE method) is used for calculating dispersion characteristics, while Ansys Lumerical FDTD is employed for time-domain simulations. The PWE method provides frequency-domain analysis, and the FDTD simulations allow for time-domain analysis, enabling a thorough investigation of the slow light effect and rainbow trapping applications.

\begin{figure}[htbp]
    \centering
    
        \includegraphics[width=\textwidth]{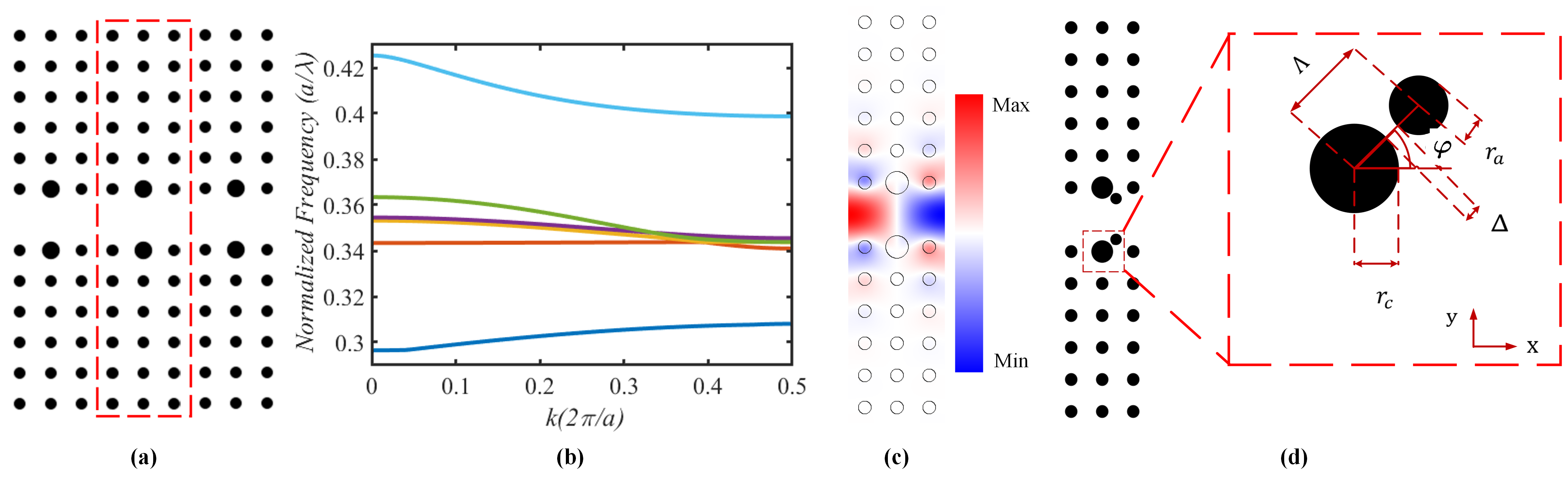}
        
    \caption{\textbf{(a)} The schematic view of the PC CCW structure with dielectric rods having radii of $r=0.2a$ and the dielectric constant of $\epsilon=9.8$. Here the enlarged rods having radii of $r_c=0.35a$ act as cavity regions, where $a$ is the lattice constant. \textbf{(b)} Multi-band TM polarization defect modes of the cavity structure without auxiliary rods in the supercell structure. \textbf{(c)} TM polarized $E_z$ Mod profile picture for $3^{rd}$ band mode with TM polarization. \textbf{(d)} The supercell of low-symmetry cavity structure with introduced auxiliary rod. Cavity parameters; $r_c$ and $r_a$ are the cavity and the auxiliary rod radii, respectively, $\varphi$ and $\Delta$ are the rotation angle of the auxiliary rod and the distance between the cavity and the auxiliary rod, respectively.}
    
    \label{fig:combined}
\end{figure}

The photonic band diagrams of both primitive and supercell PC structures are calculated using the PWE method, and by introducing line and cavity defects, we generate the desired guided modes. Fig. \ref{fig:combined}a presents the schematic view of the PC coupled cavity waveguide (CCW) structure. This structure comprises square lattice dielectric rods with a radius of $0.20a$ distributed in an air medium. The distance between the centers of adjacent rods is set to the lattice constant $a$. We design a cavity waveguide structure by integrating dielectric rods with a radius of $0.35a$ into the W1 PC waveguide. This configuration gives rise to prominent defect modes within the first band gap of the photonic band structure, as shown in Fig. \ref{fig:combined}b. The calculated mode profile of the gap-guided transverse magnetic (TM) mode at $k=0.25 (2\pi/a)$ is also presented in Fig. \ref{fig:combined}c.

%%%%fig1

\FloatBarrier

\begin{figure}[htbp]
    \centering
    
        \includegraphics[width=\textwidth]{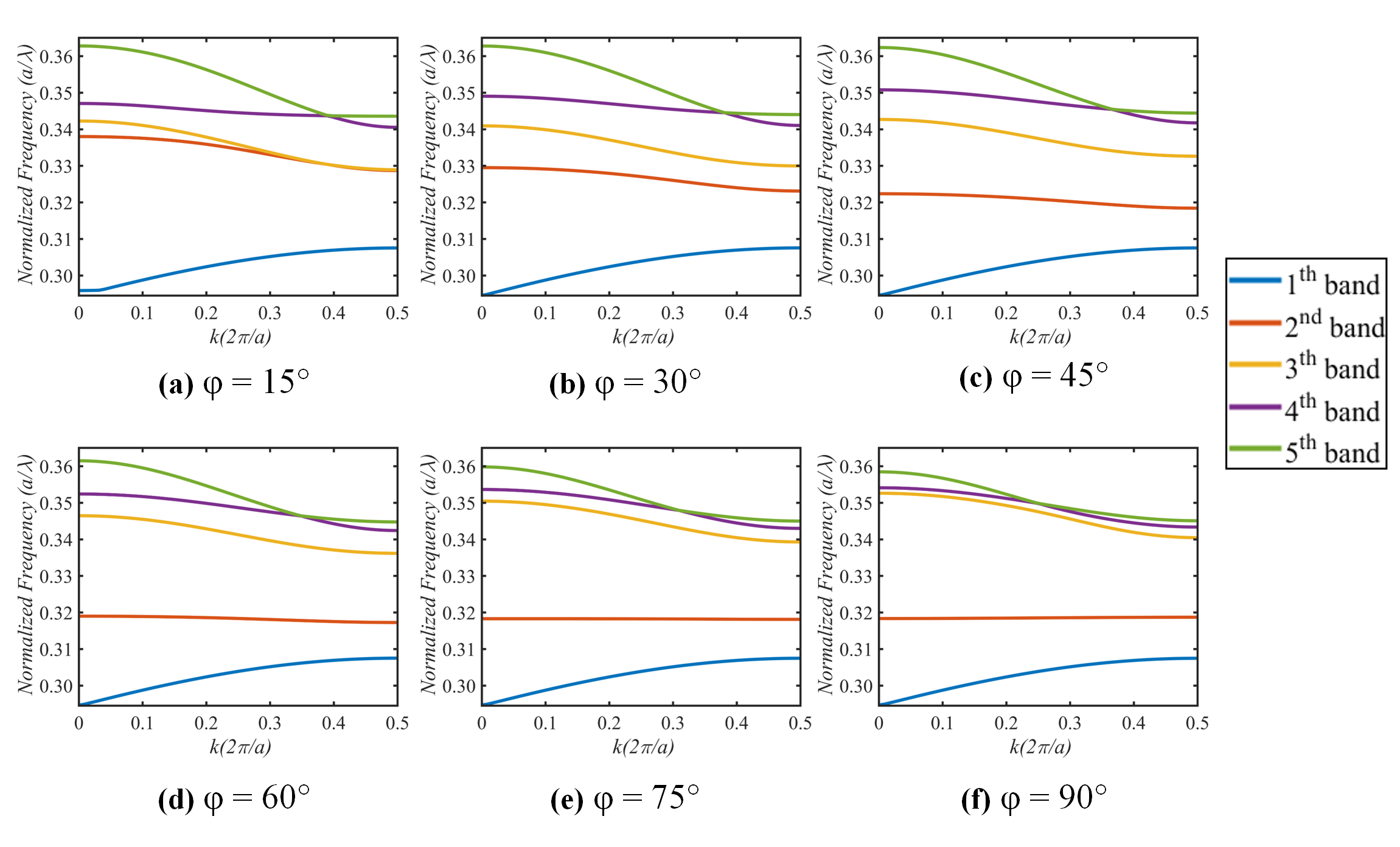}
        
    \caption{The band structure of low-symmetry cavity structure with the auxiliary rods of radius $r_a=0.14a$ at the angles of \textbf{(a)} $\varphi=15^\circ$, \textbf{(b)} $\varphi=30^\circ$, \textbf{(c)} $\varphi=45^\circ$, \textbf{(d)} $\varphi=60^\circ$, \textbf{(e)}  $\varphi=75^\circ$ and \textbf{(f)}  $\varphi=90^\circ$. The radius of cavity and the distance between the cavity and the auxiliary rod are taken as $r_c=0.35a$ and $\Delta=0.10a$, respectively.}
        \label{fig: anglesweep}       
\end{figure}

We propose a CCW design by replicating a $3a$ wide cavity supercell across the W1 waveguide and introducing symmetry reduction via auxiliary rod, as seen in Fig. \ref{fig:combined}d. The supercell of the low-symmetry cavity structure, with the introduced auxiliary rods and corresponding opto-geometrical parameters, is presented as an inset in the same figure plot. As shown in Fig. \ref{fig:combined}d, symmetry reduction is introduced in the cavity region to analyze the spectral effects of locally generated symmetry breaking on the performance of SL effect. In this regard, the band structures of symmetry-reduced PC CCW are calculated and presented in Fig. \ref{fig: anglesweep} for specifically selected auxiliary rods radius of $r_a=0.14a$ and rotation angles of auxiliary rods of  $\varphi =[ 15^\circ,  30^\circ,45^\circ,  60^\circ, 75^\circ, 90^\circ]$ as shown in Figs. \ref{fig: anglesweep}a-\ref{fig: anglesweep}f, respectively. Additionally, the radius of the cavity and the distance between the cavity and the auxiliary rod are set to $r_c = 0.35a$ and $\Delta = 0.10a$, respectively. 

The band diagrams in Figs. \ref{fig: anglesweep}a-\ref{fig: anglesweep}f, illustrate the critical effect of rotation angle $\varphi$ on the band structure. As $\varphi$ increases from $15^\circ$ to $90^\circ$, the $2^{nd}$ band shifts  to the lower frequencies, while the $3^{rd}$ and $4^{th}$ bands move closer to the $5^{th}$ band. At $\varphi =90^{\circ}$, all three bands are closely located around normalized frequencies of $a/\lambda=0.35$ and $a/\lambda=0.36$. This indicates that the rotation of the auxiliary rods from $15^\circ$ to $90^\circ$ increases the effective index $n_{eff}$ of the CCW in the propagation direction. The $n_{eff}$ characterizes the guided modes in the waveguide, representing an equivalent refractive index that accounts for the confinement and propagation characteristics within the waveguide \cite{Liu2005}. This change in $n_{eff}$ is prominent in the $2^{nd}$ band until $\varphi = 60^\circ$ and remains constant up to $90^\circ$. However, the $3^{rd}$ band shows a clear separation from the $2^{nd}$ band and exhibits unique dispersion characteristics. The $3^{rd}$ band decreases in frequency as $\varphi$ increases from $15^\circ$ to $60^\circ$, followed by an increase from $60^\circ$ to $90^\circ$. As described in Figs. \ref{fig: anglesweep}a-\ref{fig: anglesweep}d, the rotation angle of the auxiliary rods significantly alters the frequency bands of interest, paving the way for optical applications where fine-tuning the frequency response of the PC CCW is essential.

%%%%%fig 2

\FloatBarrier

\begin{figure}[htbp]
    \centering
    
        \includegraphics[width=\textwidth]{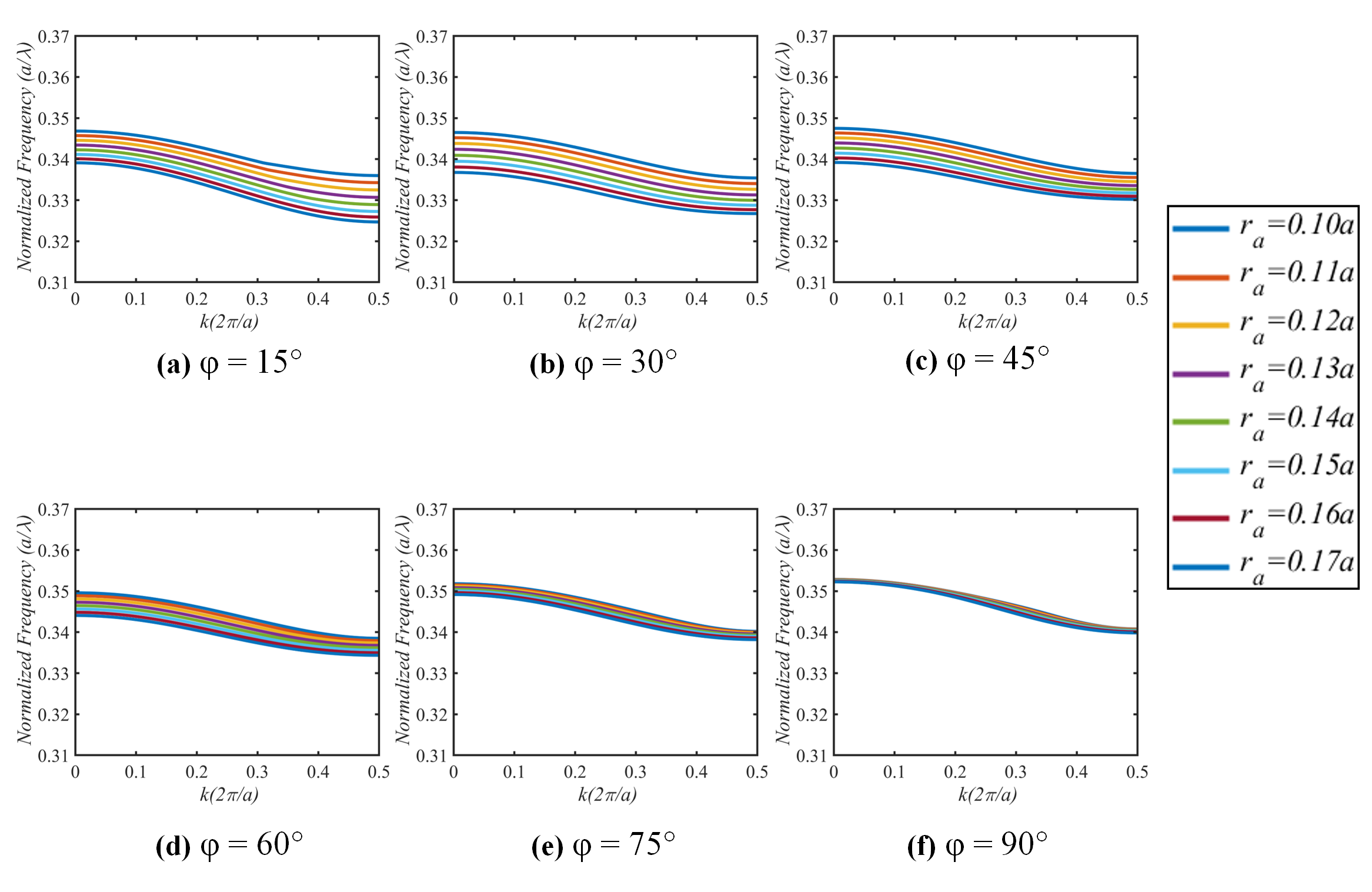}
        
 \caption{The variation of normalized frequency of the $3^{rd}$ band with the radius of auxiliary rod ranging between $r_a=0.10a-0.17a$ for the auxiliary rod angles of \textbf{(a)} $\varphi=15^\circ$, \textbf{(b)} $\varphi=30^\circ$, \textbf{(c)} $\varphi=45^\circ$, \textbf{(d)} $\varphi=60^\circ$, \textbf{(e)}  $\varphi=75^\circ$ and \textbf{(f)}  $\varphi=90^\circ$. The radius of cavity and the distance between cavity and auxiliary rod are taken as $r_c=0.35a$ and $\Delta=0.10a$, respectively.}
        \label{fig:auxsweep}       
\end{figure}

While the $2^{nd}$ band shows a flat profile, the response to symmetry reduction is limited, remaining unaffected by angle alterations between $60^\circ$ and $90^\circ$. Additionally, the $4^{th}$ and $5^{th}$ bands are closely intertwined, showing no clear separation. This makes the $3^{rd}$ band the most suitable candidate for exploitation. As seen in Figs. \ref{fig: anglesweep}a-\ref{fig: anglesweep}d, while the radius of the auxiliary rod is kept constant, the rotation angle is varied. It is also important to assess the effect of auxiliary rod radii on the frequency band of interest. In Fig. \ref{fig:auxsweep}, the impact of varying the auxiliary rod radius on the evolution of the frequency band is presented. By tuning the radius of the auxiliary rod $r_a$ from $0.10a$ to $0.17a$ and adjusting the rotation angle from $15^\circ$ to $90^\circ$, one can observe how the $3^{rd}$ frequency band is maintained, as shown in Figs. \ref{fig:auxsweep}a-\ref{fig:auxsweep}f. The separation distance is fixed at $\Delta=0.10a$. The $3^{rd}$ band exhibits a positive correlation between $r_a$ and frequency. However, as $\varphi$ increases, this effect diminishes, and the influence of $r_a$ on frequency becomes negligible after $\varphi = 75^\circ$.

%%%%%fig 3

\FloatBarrier

Notably, the alteration in $r_a$ provides a basis for refining the response characteristics of the structure. Additionally, we found that when the symmetry of the structure is reduced, angle adjustments can serve as an effective mechanism to tune the SL response, introducing another dimension of control in manipulating the SL properties of the system. This approach allows for precise control of SL parameters, such as $n_g$ and GBP, with the desired frequency responses. As shown in Figs. \ref{fig: anglesweep} and \ref{fig:auxsweep}, frequency bands exhibiting the SL effect and the effective dynamics of the targeted band modes can be achieved by introducing auxiliary rods with carefully chosen opto-geometrical parameters into the PC CCW. Based on the results presented in Fig. \ref{fig:auxsweep}, the most effective auxiliary rod radius was determined to be $r_a = 0.14a$, considering the geometrical and fabrication limits of the proposed PC CCW.

\begin{figure}[htbp]
     \centering
     \begin{subfigure}[t]{0.32\textwidth}
         \centering
         \includegraphics[width=\textwidth]{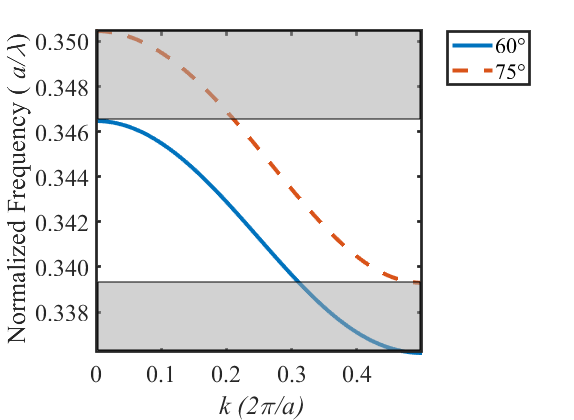}
         \caption{}
         \label{fig:6075band}
     \end{subfigure}
     \hfill
     \begin{subfigure}[t]{0.32\textwidth}
         \centering
         \includegraphics[width=\textwidth]{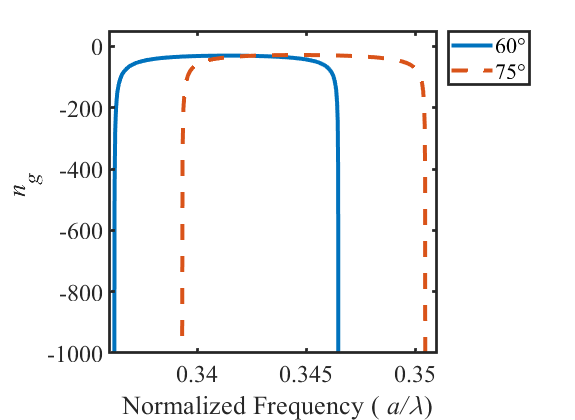}
         \caption{}
         \label{fig:6075ng}
     \end{subfigure}
     \begin{subfigure}[t]{0.32\textwidth}
         \centering
         \includegraphics[width=\textwidth]{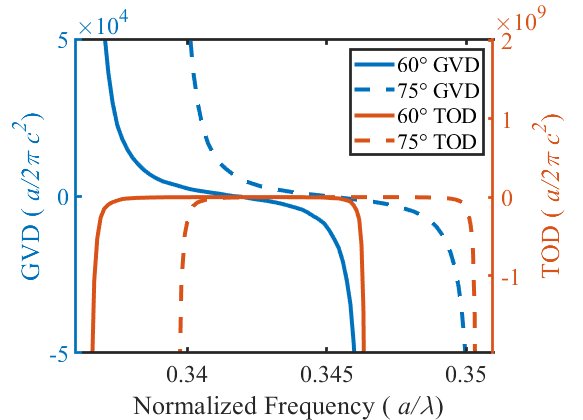}
         \caption{}
         \label{fig:6075gvdtod}
     \end{subfigure}
 \caption{\textbf{(a)} Band structure and \textbf{(b)} $n_g$, \textbf{(c)} GVD and TOD values of $3^{rd}$ band for auxillary rod angles $\varphi$ of $60^\circ$ and $75^\circ$ structures showing mini-band gap.}
        \label{fig:rt6075}       
   \end{figure}

The $1^{st}$ band is independent of symmetry reduction, as is the $2^{nd}$ band, although the latter exhibits SL characteristics, as mentioned earlier. There is a noticeable inclination in the $\varphi$ angle beyond $60^\circ$. The $4^{th}$ and $5^{th}$ bands are intertwined, and this overlap increases with the $\varphi$ angle, especially after $60^\circ$. Consequently, the $3^{rd}$ band emerges as a promising operational candidate, offering the necessary bandwidth and SL characteristics. The band structure of the $3^{rd}$ band for $\varphi$ angles of $60^\circ$ and $75^\circ$ is shown in Fig. \ref{fig:rt6075}a.

To quantitatively analyze the SL effect, we exploited the slope information of the $3^{rd}$ band mode to calculate the group index $n_g$ and group velocity $v_g$, which are presented in Fig. \ref{fig:rt6075}b for rotation angles of $\varphi = 60^\circ$ and $\varphi = 75^\circ$. Here, the relationship $v_g = c/n_g = d\omega/dk$ is employed, where $c$ represents the speed of light in a vacuum, $\omega$ is the angular frequency as normalized frequency $\omega(a/\lambda)$, and $k(2\pi/a)$ stands for the wave number.

Our research primarily focuses on conducting a detailed exploration of the impacts of symmetry reduction on the dispersion curves of waveguides. The main objective is to enhance the understanding of low-dispersion SL, characterized by low group velocity dispersion (GVD) \cite{Kubo2007}. This investigation utilizes a dual computational approach, incorporating both the PWE method and the FDTD method for frequency and time domain analysis of the PC structures, respectively \cite{mpb,taf05,meep}. The combination of these methods provides a comprehensive exploration of the influence of symmetry reduction on the properties of SL in both time and spectral domains. We calculated the photonic band diagrams of both primitive and supercell PC structures using the PWE method. By introducing line and cavity defects into the PC structure, we revealed the desired defect modes. From these defect modes, we derived the group index $n_g$ and group velocity $v_g$, which are key parameters for understanding the light propagation characteristics of the model structure.

It is also crucial to analyze SL effect in terms of group velocity dispersion (GVD) and third-order dispersion (TOD) parameters. The GVD and TOD are the key parameters used to characterize SL phenomenon in the frequency domain quantitatively. GVD, characterized by $\partial^2 \omega/\partial k^2$, dictates the propagation speeds of various frequency components of a light pulse through the medium. This is essential for optimizing the characteristics of SL, making it crucial for broadening the operational bandwidth \cite{Poon2004, Baba2008, Baba2009,Yuksel_2024}. On the other hand, TOD, represented by $\partial^3 \omega/\partial k^3$, adds another layer of complexity as it influences the temporal shape of the propagating pulse. In dispersive media, TOD becomes significant, leading to either pulse broadening or compression depending on its sign and magnitude \cite{Poon2004, Engelen2006,Baba2009}. Another useful parameter for determining how effectively different frequency components of a light signal are spatially separated and slowed down is the group index bandwidth product (GBP), which can be formulated as $GBP = \langle n_g \rangle \frac{\Delta\omega}{\omega_0}$, where $\omega_0$ is the midpoint frequency, $\Delta\omega$ is the bandwidth of the band, and $\langle n_g \rangle$ is the average group index, given by $\langle n_g \rangle = \frac{1}{\Delta\omega}\int_{\omega_0}^{\omega_0+\Delta\omega} n_g(\omega)d\omega$ \cite{Hao2010}. High GBP values allow a wider range of frequencies to be effectively trapped with significant delay, facilitating advanced optical manipulations and signal processing techniques \cite{Joannopoulos2008,Hu2013}. The corresponding GVD and TOD values are presented in Fig. \ref{fig:rt6075}c for rotation angles of $\varphi=60^\circ$ and $\varphi=75^\circ$.

%%%%fig 4

\FloatBarrier

From Fig. \ref{fig:rt6075}a, the structure creates a clear gap and a converging region, which can be effectively utilized for rainbow trapping. As discussed earlier, the frequency is closely tied to the $\varphi$ angle. When reducing symmetry for the $3^{rd}$ band, there is an increase in frequency without disturbing the bandwidth and SL parameters, as shown in Table \ref{tab:tableall} and also observed in Fig. \ref{fig:rt6075}b. Interestingly, this stability extends to TOD and GVD, as seen in Fig. \ref{fig:rt6075}c, ensuring minimal pulse deformation while breaking symmetry. This characteristic can be particularly useful for designing RT-based devices.

\section{Symmetry Breaking and Rainbow Trapping}

While low-symmetry applied to the overall structure has been reported in the literature \cite{He2021}, the focus of our study is to investigate the effects of locally generated symmetry reduction in the cavity, specifically achieved by integrating an auxiliary rod at various angles to the main cavity rod. We aim to leverage this approach to control SL, GBP, and transmission parameters. Our design adopts a square lattice configuration with a lattice constant $a$ and rods of radius $r=0.2a$, as shown in Fig. \ref{fig:combined}. This structure features a W1 waveguide created by removing a single row of rods. The supercell of the design is $3a$ wide. A cavity is formed in the supercell by modulating the rod radii of the mirror-symmetric middle rods to $r_c=0.35a$. This design yields a primary bandgap in the normalized frequency range of $(0.30-0.40) a/\lambda$, with multiband guided modes appearing approximately within the range of $(0.34-0.37) a/\lambda$. We focus on achieving separation by adding low-symmetry alterations for RT, which produces defect modes exhibiting grouped behavior and degeneration, as shown in Fig. \ref{fig:combined}. The insights garnered from this study are expected to advance the understanding of structural effects on photonic characteristics, particularly in contexts requiring fine control over SL and GBP parameters with minimal symmetry perturbations.

Our study progresses by conducting PWE-based simulations to examine the effects of the asymmetrical angle of auxiliary rods on the optical properties of the RT model. We calculate the band structure and SL parameters, including $n_g$, $v_g$, GBP, GVD, and TOD. The structure of the supercell and the structural parameters of the RT model are shown in Fig. \ref{fig:combined}d. The SL parameters exhibit considerable fluctuations depending on the angle of the auxiliary rod. Notably, at certain angles, the group index is maximized while maintaining a relatively high GBP in the SL region. We then leverage the symmetry-breaking effect by combining two different auxiliary rod angles within the cavity to achieve RT with SL effect, creating mini-band gaps that can be effectively used for RT. As the angle of the auxiliary rods increases, the band frequency decreases, which is directly related to the symmetry reduction, as shown in Figs. 2a-2f. Specifically, as the angle between the cavity (located along the \textit{x-axis}) and the auxiliary rod increases from $15^\circ$ to $60^\circ$, both the cutoff frequency and center frequency $\omega_0$ decrease significantly. This behavior has the potential to fine-tune the frequency response of the proposed structure.

The effect of the auxiliary rod angle on the band structure of the PC is most clearly observed in the first SL defect mode, where the auxiliary rod radius is $r_a = 0.14a$. This phenomenon manifests as a distinct separation of band modes, reaching a maximum at an angle of $90^\circ$, as depicted in Fig. \ref{fig:auxsweep} . By modulating the radius $r_a$ of the auxiliary rod within the range of $0.10a-0.17a$ while keeping the separation distance constant at $\Delta = 0.10a$, we successfully improved the frequency response of the $3^{rd}$ band, as shown in Figs. 3a-3f. Notably, this adjustment in $r_a$ provides a foundation for refining the response characteristics of the structure. Additionally, we found that when the symmetry of the structure is reduced, angle adjustments can serve as a mechanism to tune the SL response, introducing an additional dimension of control over the SL properties of the system. In this way, we can control SL parameters such as $n_g$ and GBP to achieve the desired frequency responses. Using these techniques, we achieved RT by combining different angle CCWs through symmetry breaking. A conspicuous separation of the band modes is particularly evident when the auxiliary rod radius is $r_a = 0.14a$. This separation forms the basis of our RT efforts. As seen in Fig. \ref{fig:6075band}, this separation is most prominent around the $60^\circ$ and $75^\circ$ auxiliary rod angles, leading to a marked band gap that can be exploited for RT. Moreover, as seen in Fig. \ref{fig:6075ng}, the SL effects are promising for any SL-based applications. Additionally, the relatively low GVD and TOD behaviors of the band, as observed in Fig. \ref{fig:6075gvdtod}, may be beneficial for light manipulation without disturbing the propagating wave shape, making them advantageous for RT applications.

Table \ref{tab:tableall} summarizes the results of GBP for the structure with $r_a=0.14a$ auxiliary rods at the angles of $\varphi=60^\circ$ and $\varphi=75^\circ$ (Model B), along with a comparison of the results for the structure without auxiliary rods (Model A). The total trapping performance characteristics of the structure can be easily obtained from Fig. \ref{fig:6075band}. For the $3^{rd}$ band modes, the bandwidth and the midpoint frequency for trapping are calculated as $\Delta\omega=0.0061 a/\lambda$ and $\omega_0=0.3380 a/\lambda$, respectively. Table \ref{tab:tableall} also includes information about the SL performance. Notably, there are significant increases in both $\Delta\omega$ and $\langle n_g \rangle$ when comparing Model A to Model B. Model B, with $\varphi=60^\circ$ and $\varphi=75^\circ$ angle values, exhibits a $50.07 \%$ and $32.67 \%$ increase in $\Delta\omega$, respectively. Similarly, the $\langle n_g \rangle$ values in Model B increase by $282.16 \%$ for $\varphi=60^\circ$ and $1813.72 \%$ for $\varphi=75^\circ$. The substantial differences in parameters for $\varphi = 60^\circ$ and $\varphi = 75^\circ$ are primarily due to the altered field distribution and mode coupling introduced by the different rotational angles of the auxiliary rods. These angles impact the photonic band structure, leading to distinct dispersion and bandwidth characteristics. Consequently, this configuration demonstrates enhanced SL properties and is suitable for broken symmetry RT applications.

\begin{table}[H]
\centering
\resizebox{0.95\textwidth}{!}{
\begin{tabular}{@{}crrrrrrr@{}}
\toprule
Model  & $r_a$      & $\varphi$    & $\omega_0$ ($a/\lambda$)& $\Delta\omega$ ($a/\lambda$)&  $\langle n_g \rangle$  &  $\langle v_g \rangle  (c)$  & $GBP$    \\ \midrule

A & $-$    & $-$  & $0.353157000$ &$ 0.010815$ &  $162.2441584$ & $0.018918020$  &  $0.506709799$\\
B & $0.14$ & $60^\circ$ & $0.341294000$ & $0.016230$ &  $620.0413845$ & $0.020572013$  &  $3.420130191$\\
B & $0.14$ & $75^\circ$ & $0.345040969$ & $0.014351$ & $3104.8971230$ & $0.022326015$  &  $3.425385380$\\
  \bottomrule
\end{tabular} }
\caption{The midpoint frequency $\omega_0$, the bandwidth $\Delta\omega$, the average group index  $\langle n_g \rangle$, the average group velocity  $\langle v_g \rangle$ times light velocity in vacuum $c$, $GBP$ for the $3^{rd}$ band mode of PC cavity without auxiliary rods (Model A) and with auxiliary rods at the angles of  $60^\circ$ and $75^\circ$ (Model B). All the values are valid for the radius of auxiliary rods $r_a=0.14a$. }
    \label{tab:tableall}
 \end{table}
   
\FloatBarrier

%%%%fig 5
\begin{figure}[htbp]
    \centering
  
        \centering
        \begin{subfigure}[b]{0.95\textwidth}
            \includegraphics[width=\textwidth]{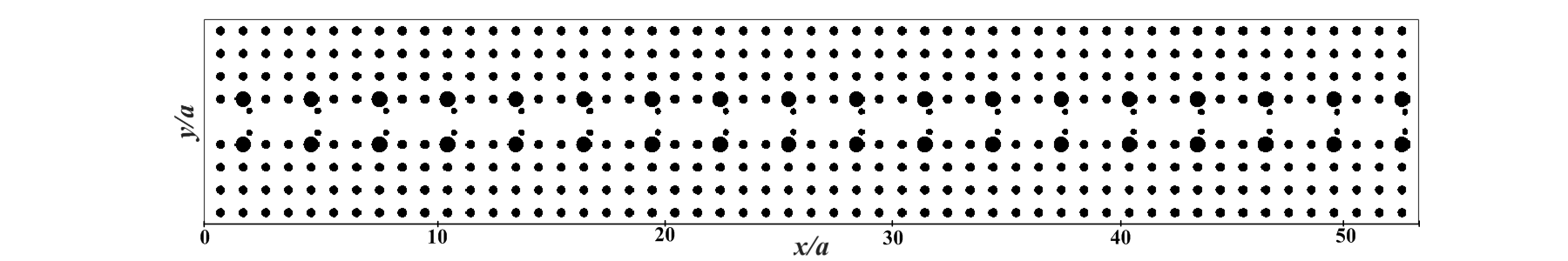}
            \caption{RT structure}
            \label{fig: rtst}
        \end{subfigure}
            \begin{subfigure}[b]{0.45\textwidth}
            \includegraphics[width=\textwidth]{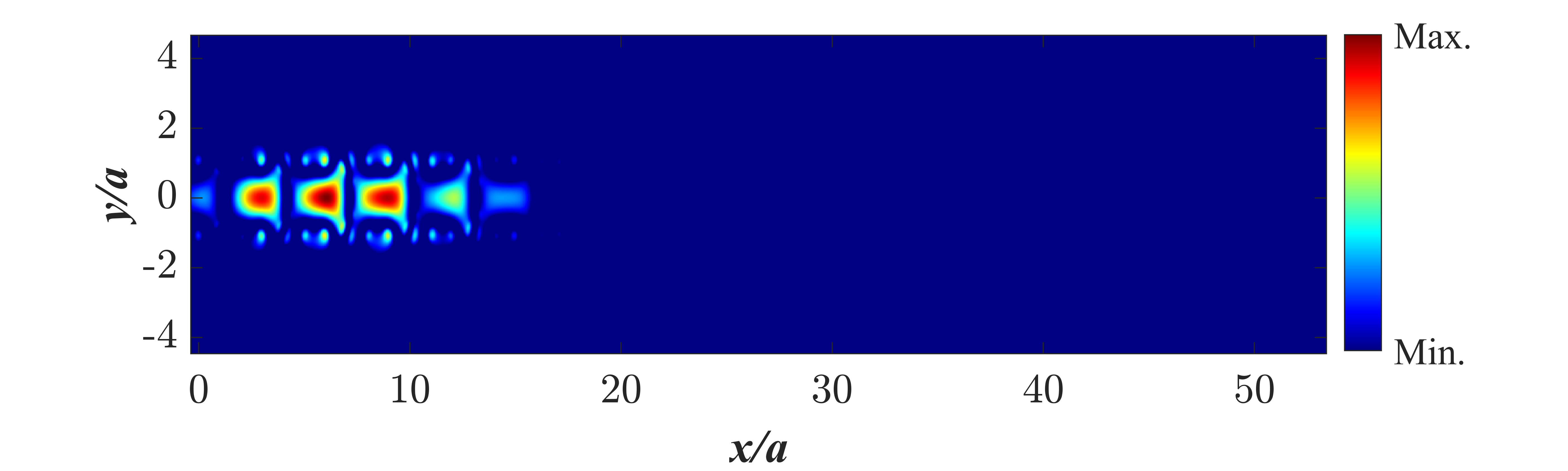}
            \caption{$0.3372 a/\lambda$}
            \label{fig: rt01}
        \end{subfigure}
        \begin{subfigure}[b]{0.45\textwidth}
            \includegraphics[width=\textwidth]{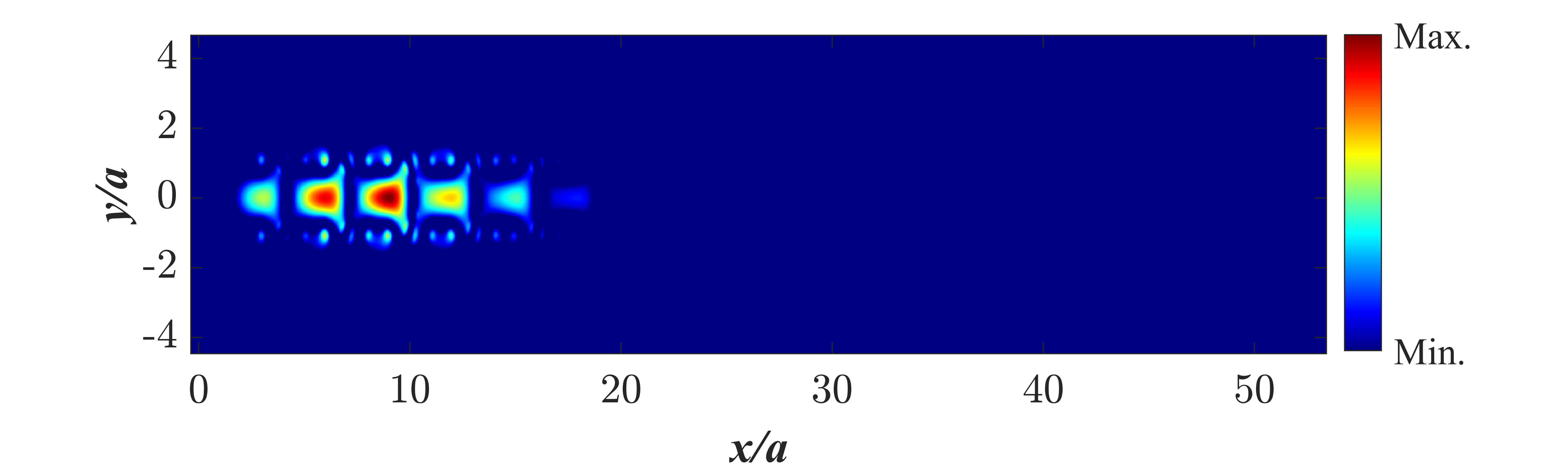}
            \caption{$0.3376 a/\lambda$}
            \label{fig: rt02}
        \end{subfigure}
        \begin{subfigure}[b]{0.45\textwidth}
            \includegraphics[width=\textwidth]{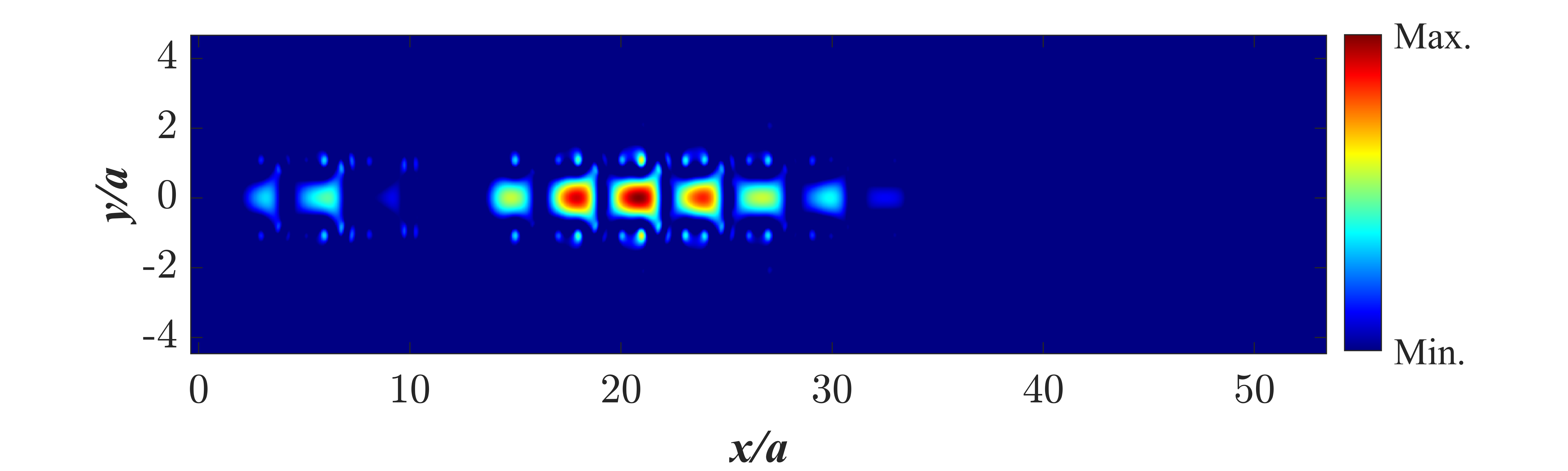}
            \caption{$0.3382 a/\lambda$}
            \label{fig: rt03}
        \end{subfigure}
                \centering
            \begin{subfigure}[b]{0.45\textwidth}
            \includegraphics[width=\textwidth]{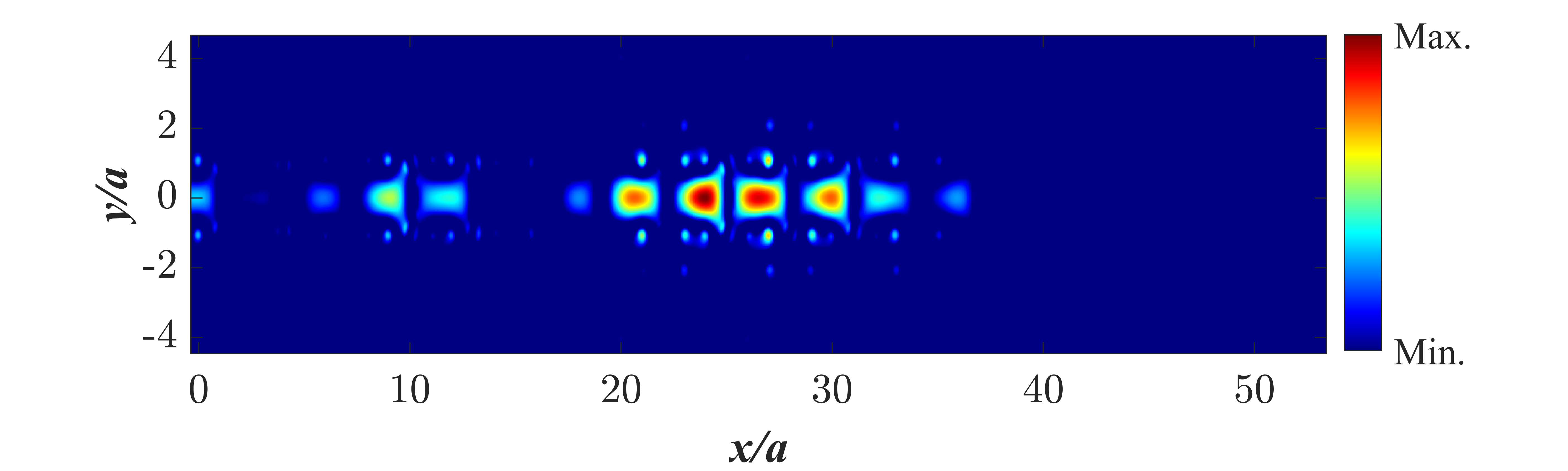}
            \caption{$0.3386 a/\lambda$}
            \label{fig: rt04}
        \end{subfigure}
        \begin{subfigure}[b]{0.45\textwidth}
            \includegraphics[width=\textwidth]{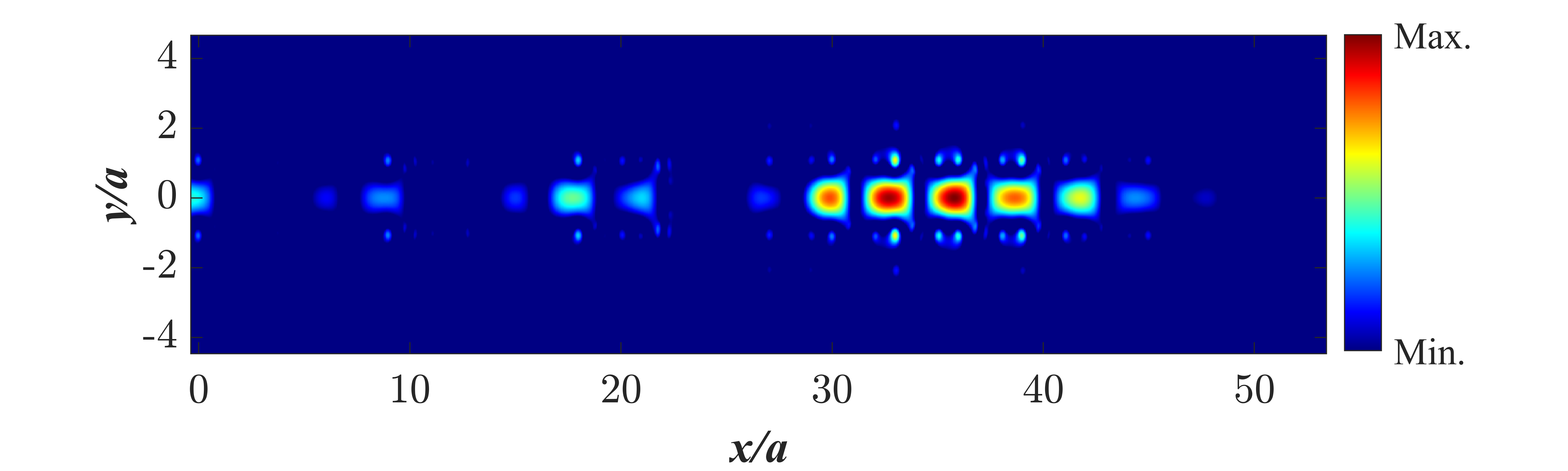}
            \caption{$0.3392 a/\lambda$}
            \label{fig: rt05}
        \end{subfigure}
        \begin{subfigure}[b]{0.45\textwidth}
            \includegraphics[width=\textwidth]{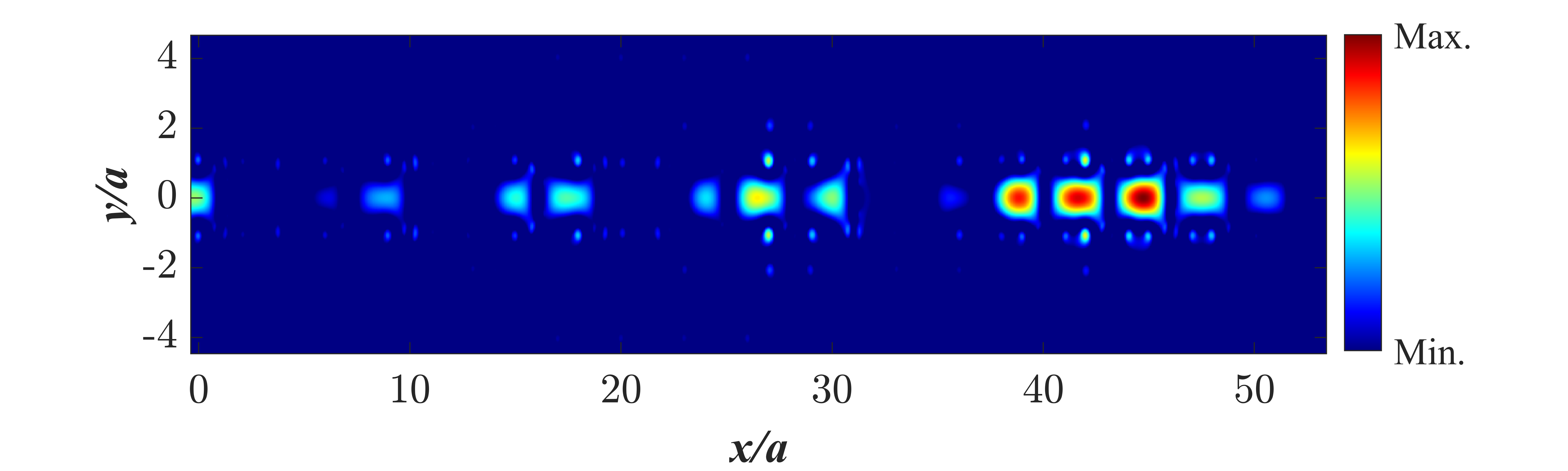}
            \caption{$0.3396 a/\lambda$}
            \label{fig: rt06}
        \end{subfigure}
           \begin{subfigure}[b]{0.7\textwidth}
            \includegraphics[width=\textwidth]{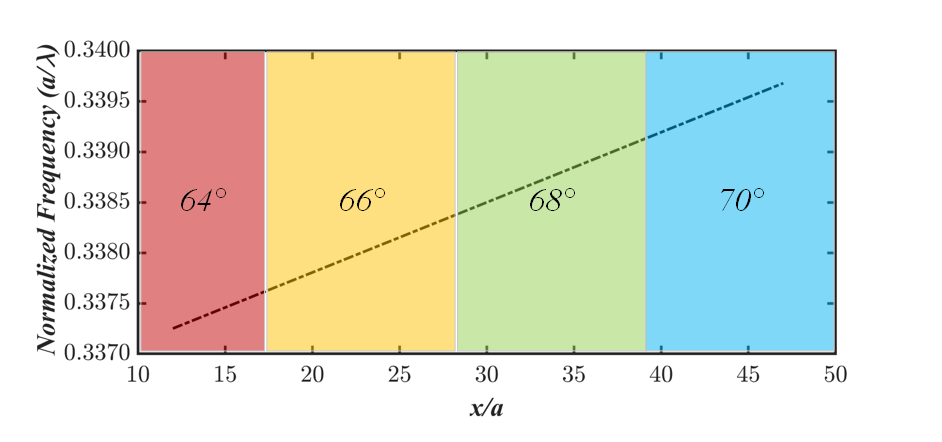}
            \caption{Spatial RT positions corresponding to frequency}
            \label{fig: rtsp}
        \end{subfigure}
        
    \hfill

    \caption{The structure schematics and simulated electric field profiles for $\varphi=60^\circ$ and $\varphi=75^\circ$ with $2^\circ$ gradual increase: \textbf{(a)} Top view of the photonic crystal coupled cavity waveguide with the starting auxiliary rods of the angle of $\varphi=60^\circ$ and gradually increasing to $\varphi=75^\circ$, $E_z$ field intensities of \textbf{(b)}  $\omega=0.3372 a/\lambda$, \textbf{(c)}  $\omega=0.3376 a/\lambda$, \textbf{(d)} $\omega=0.3382 a/\lambda$, \textbf{(e)} $\omega=0.3386 a/\lambda$, \textbf{(f)} $\omega=0.3392 a/\lambda$, \textbf{(g)} $\omega=0.3396 a/\lambda$ and \textbf{(h)}  spatial RT positions corresponding to frequency.}
    \label{fig: rtall} 
\end{figure}

\FloatBarrier

To achieve RT, we implemented the gradual increase in the $\varphi$ angle with $2^\circ$ increments, as shown in Fig. \ref{fig: rtst}. By introducing this lower symmetry alteration to the structure and gradually breaking symmetry, we successfully trapped light spatially within the normalized frequency range of $(0.3372-0.3396) a/\lambda$, with the $E_z$ field profiles depicted in Figs. \ref{fig: rt01}-\ref{fig: rt06}. These field profiles demonstrate that the localization is closely tied to frequency. The mini-band gap localization is also linked to the $\varphi$ angle. Small structural alterations create a shifting mini-band gap, resulting in spatial localizations. The spatial localization of these frequencies is clearly shown in Fig. \ref{fig: rtsp}, where the linear relationship between trapping position and frequency is evident. This structure is suitable for RT applications and can be implemented in any RT-based devices.

It is important to note that in the presented study, our primary objective was to demonstrate the concept of rainbow trapping through symmetry breaking in the waveguide defect region, rather than to fully develop and test the demultiplexing application. We conducted several simulations to demonstrate lateral light coupling at selected operating frequencies. While lateral channels were introduced at locations corresponding to light trapping, we acknowledge that these coupling mechanisms require further optimization for efficiency. It is important to clarify that the application of demultiplexing was beyond the scope of this initial study. Our results primarily aim to establish the foundational concept, paving the way for future studies to explore and optimize demultiplexing configurations in detail. Additionally, the robustness of the PC structure against imperfections cannot be overlooked, as it could indeed affect the practical application of the observed effects. In our simulations, we introduced variations to test the structure's sensitivity to imperfections. We found that $a \pm\%5$ alteration in the radii of the dielectric rods and the spacing between them can cause shifts in the operating frequencies. Although these shifts can impact the dispersion properties, the fundamental concept of the light-trapping effect remains intact. This suggests that while exact precision is crucial for specific applications, the overall phenomenon of rainbow trapping can be robust under minor variations.

\section{Conclusion}
   
In this study, we explored the use of auxiliary rods to enhance SL parameters by introducing rotational symmetry reduction. Additionally, we achieved rainbow trapping (RT) by selectively applying low-symmetry modifications to the cavity structure in a novel way. Our technique offers clear advantages in device fabrication. The stable $\omega_0$ and $\Delta\omega$ values achieved in this study make it feasible to design all-optically controlled delay lines for timing functions. The addition of auxiliary rods significantly improves both $n_g$ and $\Delta\omega$ for the $3^{rd}$ band. For the proposed $\varphi=60^\circ$ and $\varphi=75^\circ$ structures, $\Delta\omega$ increased by $50.07 \%$ and $32.67 \%$, respectively, while $\langle n_g \rangle$ increased by $282.16 \%$ and $1813.72 \%$. These improvements notably enhance the GBP of the $3^{rd}$ band by $675 \%$, underscoring the potential of our findings. The wave propagation properties resulting from the gradual increase in broken symmetry from $\varphi=60^\circ$ to $\varphi=75^\circ$ present intriguing possibilities. The clear linear relationship between spatial RT positions and frequency may prove advantageous for any RT-based application. The entire structure exhibits relatively flat and low GVD and TOD values, offering opportunities for demultiplexer and multiplexer devices that maintain wave integrity while achieving SL effects and RT. This also opens up avenues for time-domain and spatial signal processing. For further controllability, the auxiliary rods could be manipulated by external means, such as mechanical or thermal interactions, enabling active control of light. This presents new opportunities for research in this domain and could lead to advancements in nanotechnology applications. The normalized frequency results could be adapted to various frequency bands, such as the terahertz region or the microwave domain. Overall, the use of auxiliary rods to achieve enhanced SL parameters with RT shows significant potential for the development of advanced optical devices with a wide range of applications.

\section{Acknowledgement}

This study was supported by the Scientific Research Coordination Unit of Pamukkale University under project number 2021FEBE041 and YÖK 100/2000 priority fields doctoral scholarship program. Also, we thank the anonymous referees for their guiding questions and criticisms.

\newpage
\bibliographystyle{unsrtnat}
\bibliography{bibi.bib}

\begin{thebibliography}{47}
\providecommand{\natexlab}[1]{#1}
\providecommand{\url}[1]{\texttt{#1}}
\expandafter\ifx\csname urlstyle\endcsname\relax
  \providecommand{\doi}[1]{doi: #1}\else
  \providecommand{\doi}{doi: \begingroup \urlstyle{rm}\Url}\fi

\bibitem[Yablonovitch(1993)]{Yablonovitch1993}
E.~Yablonovitch.
\newblock Photonic band-gap structures.
\newblock \emph{J. Opt. Soc. Am. B}, 10\penalty0 (2):\penalty0 283--295, Feb
  1993.
\newblock \doi{10.1364/JOSAB.10.000283}.

\bibitem[Joannopoulos et~al.(2008)Joannopoulos, Johnson, Winn, and
  Meade]{Joannopoulos2008}
J.~D. Joannopoulos, S.~G. Johnson, J.~N. Winn, and R.~D. Meade.
\newblock \emph{Photonic Crystals: Molding the Flow of Light (Second Edition)}.
\newblock Princeton University Press, 2 edition, 2008.
\newblock ISBN 0691124566.
\newblock URL \url{http://ab-initio.mit.edu/book/}.

\bibitem[Xia et~al.(2007)Xia, Rooks, Sekaric, and Vlasov]{Xia2007}
F.~Xia, M.~Rooks, L.~Sekaric, and Y.~Vlasov.
\newblock Ultra-compact high order ring resonator filters using submicron
  silicon photonic wires for on-chip optical interconnects.
\newblock \emph{Opt. Express}, 15\penalty0 (19):\penalty0 11934--11941, Sep
  2007.
\newblock \doi{10.1364/OE.15.011934}.

\bibitem[Zhang et~al.(2008)Zhang, Song, Wu, Zou, Beausoleil, and
  Willner]{Zhang2008}
L.~Zhang, M.~Song, T.~Wu, L.~Zou, R.~G. Beausoleil, and A.~E. Willner.
\newblock Embedded ring resonators for microphotonic applications.
\newblock \emph{Opt. Lett.}, 33\penalty0 (17):\penalty0 1978--1980, Sep 2008.
\newblock \doi{10.1364/OL.33.001978}.

\bibitem[Notomi et~al.(2008)Notomi, Kuramochi, and Tanabe]{notomi2008}
M.~Notomi, E.~Kuramochi, and T.~Tanabe.
\newblock Large-scale arrays of ultrahigh-q coupled nanocavities.
\newblock \emph{Nature Photonics}, 2\penalty0 (12):\penalty0 741--747, Dec
  2008.
\newblock \doi{10.1038/nphoton.2008.226}.

\bibitem[Xiao et~al.(2007)Xiao, Khan, Shen, and Qi]{Xiao2007}
S.~Xiao, M.~H. Khan, H.~Shen, and M.~Qi.
\newblock A highly compact third-order silicon microring add-drop filter with a
  very large free spectral range, a flat passband and a low delay dispersion.
\newblock \emph{Opt. Express}, 15\penalty0 (22):\penalty0 14765--14771, Oct
  2007.
\newblock \doi{10.1364/OE.15.014765}.

\bibitem[Kubo et~al.(2007)Kubo, Mori, and Baba]{Kubo2007}
S.~Kubo, D.~Mori, and T.~Baba.
\newblock Low-group-velocity and low-dispersion slow light in photonic crystal
  waveguides.
\newblock \emph{Opt. Lett.}, 32\penalty0 (20):\penalty0 2981--2983, Oct 2007.
\newblock \doi{10.1364/OL.32.002981}.
\newblock URL \url{http://opg.optica.org/ol/abstract.cfm?URI=ol-32-20-2981}.

\bibitem[Baba(2008)]{Baba2008}
T.~Baba.
\newblock Slow light in photonic crystals.
\newblock \emph{Nature Photonics}, 2\penalty0 (8):\penalty0 465--473, Aug 2008.
\newblock \doi{10.1038/nphoton.2008.146}.
\newblock URL \url{https://doi.org/10.1038/nphoton.2008.146}.

\bibitem[Shu and Mao(2015)]{Shu2015}
J.~Shu and Y.~Mao.
\newblock Study of the properties of slow light in planar photonic crystal
  coupled-cavity waveguides.
\newblock In \emph{AOPC 2015: Advances in Laser Technology and Applications},
  volume 9671, page 96710O. SPIE, 2015.
\newblock \doi{10.1117/12.2199249}.

\bibitem[Notomi et~al.(2001)Notomi, Yamada, Shinya, Takahashi, Takahashi, and
  Yokohama]{notomi2001}
M.~Notomi, K.~Yamada, A.~Shinya, J.~Takahashi, C.~Takahashi, and I.~Yokohama.
\newblock Extremely large group-velocity dispersion of line-defect waveguides
  in photonic crystal slabs.
\newblock \emph{Phys. Rev. Lett.}, 87:\penalty0 253902, Nov 2001.
\newblock \doi{10.1103/PhysRevLett.87.253902}.
\newblock URL \url{https://link.aps.org/doi/10.1103/PhysRevLett.87.253902}.

\bibitem[\"{U}st\"{u}n and Kurt(2010)]{Ustun2010}
K.~\"{U}st\"{u}n and H.~Kurt.
\newblock Ultra slow light achievement in photonic crystals by merging coupled
  cavities with waveguides.
\newblock \emph{Opt. Express}, 18\penalty0 (20):\penalty0 21155--21161, Sep
  2010.
\newblock \doi{10.1364/OE.18.021155}.

\bibitem[Povinelli et~al.(2005)Povinelli, Johnson, and
  Joannopoulos]{povinelli2005}
M.~L. Povinelli, S.~G. Johnson, and J.~D. Joannopoulos.
\newblock Slow-light, band-edge waveguides for tunable time delays.
\newblock \emph{Opt. Express}, 13\penalty0 (18):\penalty0 7145--7159, 2005.
\newblock \doi{10.1364/OPEX.13.007145}.
\newblock URL \url{http://opg.optica.org/oe/abstract.cfm?URI=oe-13-18-7145}.

\bibitem[Giden et~al.(2013)Giden, Turduev, and Kurt]{GIDEN2013}
I.~H. Giden, M.~Turduev, and H.~Kurt.
\newblock Broadband super-collimation with low-symmetric photonic crystal.
\newblock \emph{Photonics and Nanostructures - Fundamentals and Applications},
  11\penalty0 (2):\penalty0 132--138, 2013.
\newblock \doi{https://doi.org/10.1016/j.photonics.2012.12.001}.
\newblock URL
  \url{https://www.sciencedirect.com/science/article/pii/S1569441012001216}.

\bibitem[Li et~al.(2008)Li, White, O'Faolain, Gomez-Iglesias, and
  Krauss]{Li2008}
J.~Li, T.~P. White, L.~O'Faolain, A.~Gomez-Iglesias, and T.~F. Krauss.
\newblock Systematic design of flat band slow light in photonic crystal
  waveguides.
\newblock \emph{Opt. Express}, 16\penalty0 (9):\penalty0 6227--6232, Apr 2008.
\newblock \doi{10.1364/OE.16.006227}.
\newblock URL \url{http://opg.optica.org/oe/abstract.cfm?URI=oe-16-9-6227}.

\bibitem[Moghaddam et~al.(2013)Moghaddam, Attari, and Mirsalehi]{khatibi2013}
M.~Khatibi Moghaddam, A.~R. Attari, and M.~M. Mirsalehi.
\newblock High coupling efficiency to a low dispersion slow light-supporting
  photonic crystal waveguide.
\newblock \emph{Journal of the European Optical Society - Rapid publications},
  8\penalty0 (0), 2013.
\newblock ISSN 1990-2573.
\newblock URL \url{https://www.jeos.org/index.php/jeos_rp/article/view/13066}.

\bibitem[Hao et~al.(2010)Hao, Cassan, Kurt, Roux, Marris-Morini, Vivien, Wu,
  Zhou, and Zhang]{Hao2010}
R.~Hao, E.~Cassan, H.~Kurt, X.~Le Roux, D.~Marris-Morini, L.~Vivien, H.~Wu,
  Z.~Zhou, and X.~Zhang.
\newblock Novel slow light waveguide with controllable delay-bandwidth product
  and utra-low dispersion.
\newblock \emph{Opt. Express}, 18\penalty0 (6):\penalty0 5942--5950, Mar 2010.
\newblock \doi{10.1364/OE.18.005942}.
\newblock URL \url{http://opg.optica.org/oe/abstract.cfm?URI=oe-18-6-5942}.

\bibitem[Bagci and Akaoglu(2015)]{Bagci2015}
F.~Bagci and B.~Akaoglu.
\newblock Enhancement of buffer capability in slow light photonic crystal
  waveguides with extended lattice constants.
\newblock \emph{Optical and Quantum Electronics}, 47\penalty0 (3):\penalty0
  791--806, Mar 2015.
\newblock \doi{10.1007/s11082-014-9953-8}.
\newblock URL \url{https://doi.org/10.1007/s11082-014-9953-8}.

\bibitem[Danaie et~al.(2018)Danaie, Geravand, and Mohammadi]{Danaie2018}
M.~Danaie, A.~Geravand, and S.~Mohammadi.
\newblock Photonic crystal double-coupled cavity waveguides and their
  application in design of slow-light delay lines.
\newblock \emph{Photonics and Nanostructures - Fundamentals and Applications},
  28:\penalty0 61--69, 2018.
\newblock \doi{https://doi.org/10.1016/j.photonics.2017.11.009}.
\newblock URL
  \url{https://www.sciencedirect.com/science/article/pii/S1569441017302808}.

\bibitem[Schulz et~al.(2010)Schulz, O'Faolain, Beggs, White, Melloni, and
  Krauss]{Schulz2010}
S.~A. Schulz, L.~O'Faolain, D.~M. Beggs, T.~P. White, A.~Melloni, and T.~F.
  Krauss.
\newblock Dispersion engineered slow light in photonic crystals: a comparison.
\newblock \emph{Journal of Optics}, 12\penalty0 (10):\penalty0 104004, Sep
  2010.
\newblock \doi{10.1088/2040-8978/12/10/104004}.

\bibitem[He et~al.(2018)He, Gao, Jiang, Wang, Zhou, and Xu]{he2018}
L.~He, Y.~F. Gao, Z.~Jiang, L.~S. Wang, J.~Zhou, and X.~F. Xu.
\newblock A unidirectional air waveguide basing on coupling of two self-guiding
  edge modes.
\newblock \emph{Optics \& Laser Technology}, 108:\penalty0 265--272, 2018.
\newblock ISSN 0030-3992.
\newblock \doi{https://doi.org/10.1016/j.optlastec.2018.06.044}.
\newblock URL
  \url{https://www.sciencedirect.com/science/article/pii/S0030399218306790}.

\bibitem[Moghaddam and Fleury(2019)]{Moghaddam2019}
M.~K. Moghaddam and R.~Fleury.
\newblock Slow light engineering in resonant photonic crystal line-defect
  waveguides.
\newblock \emph{Opt. Express}, 27\penalty0 (18):\penalty0 26229--26238, Sep
  2019.
\newblock \doi{10.1364/OE.27.026229}.
\newblock URL \url{https://opg.optica.org/oe/abstract.cfm?URI=oe-27-18-26229}.

\bibitem[Shanhui and Joannopoulos(2002)]{Shanhui2002}
F.~Shanhui and J.~D. Joannopoulos.
\newblock Analysis of guided resonances in photonic crystal slabs.
\newblock \emph{Phys. Rev. B}, 65:\penalty0 235112, Jun 2002.
\newblock \doi{10.1103/PhysRevB.65.235112}.
\newblock URL \url{https://link.aps.org/doi/10.1103/PhysRevB.65.235112}.

\bibitem[Yasa et~al.(2017)Yasa, Turduev, Giden, and Kurt]{Yasa2017}
U.~G. Yasa, M.~Turduev, I.~H. Giden, and H.~Kurt.
\newblock High extinction ratio polarization beam splitter design by
  low-symmetric photonic crystals.
\newblock \emph{J. Lightwave Technol.}, 35\penalty0 (9):\penalty0 1677--1683,
  May 2017.
\newblock URL \url{http://opg.optica.org/jlt/abstract.cfm?URI=jlt-35-9-1677}.

\bibitem[Gumus et~al.(2019)Gumus, Tutgun, Yilmaz, and Kurt]{Gumus2019}
M.~A. Gumus, M.~Tutgun, D.~Yilmaz, and H.~Kurt.
\newblock A reduced symmetric 2d photonic crystal cavity with wavelength
  tunability.
\newblock \emph{Journal of Physics D: Applied Physics}, 52\penalty0
  (32):\penalty0 325103, Jun 2019.
\newblock \doi{10.1088/1361-6463/ab1f4b}.
\newblock URL \url{https://doi.org/10.1088/1361-6463/ab1f4b}.

\bibitem[Erim et~al.(2019)Erim, Erim, and Kurt]{Erim2019}
N.~Erim, M.~N. Erim, and H.~Kurt.
\newblock An optical sensor design using surface modes of low-symmetric
  photonic crystals.
\newblock \emph{IEEE Sensors Journal}, 19\penalty0 (14):\penalty0 5566--5573,
  Jul 2019.
\newblock \doi{10.1109/JSEN.2019.2909863}.
\newblock URL \url{https://doi.org/10.1109/JSEN.2019.2909863}.

\bibitem[He et~al.(2021)He, Wu, Feng, Su, and Li]{He2021}
C.~He, H.~Wu, Y.~Feng, W.~Su, and F.~Li.
\newblock Dynamic modulation of slow light rainbow trapping and releasing in a
  tapered waveguide based on low-symmetric photonic crystals.
\newblock \emph{Results in Physics}, 28:\penalty0 104592, 2021.
\newblock \doi{10.1016/j.rinp.2021.104592}.
\newblock URL
  \url{https://www.sciencedirect.com/science/article/pii/S2211379721006896}.

\bibitem[Vucic et~al.(2001)Vucic, Loncar, Mabuchi, and Scherer]{Vucic2001}
J.~Vucic, M.~Loncar, H.~Mabuchi, and A.~Scherer.
\newblock Design of photonic crystal microcavities for cavity qed.
\newblock \emph{Phys. Rev. E}, 65:\penalty0 016608, Dec 2001.
\newblock \doi{10.1103/PhysRevE.65.016608}.
\newblock URL \url{https://link.aps.org/doi/10.1103/PhysRevE.65.016608}.

\bibitem[Gao et~al.(2020)Gao, He, Jiang, Zhou, Shi, and Bai]{Gao2020}
Y.~Gao, L.~He, Z.~Jiang, J.~Zhou, Y.~Shi, and W.~Bai.
\newblock Investigation of coupling effect between a unidirectional air
  waveguide and two cavities with one-way rotating state.
\newblock \emph{Optica Applicata}, Vol. 50, nr 1:\penalty0 49--59, 2020.
\newblock \doi{10.37190/oa200104}.

\bibitem[Yang et~al.(2015)Yang, Zhu, and Li]{Yang2015}
R.~Yang, W.~Zhu, and J.~Li.
\newblock Realization of ``trapped rainbow'' in 1d slab waveguide with surface
  dispersion engineering.
\newblock \emph{Opt. Express}, 23\penalty0 (5):\penalty0 6326--6335, Mar 2015.
\newblock \doi{10.1364/OE.23.006326}.
\newblock URL \url{https://opg.optica.org/oe/abstract.cfm?URI=oe-23-5-6326}.

\bibitem[Gao et~al.(2012)Gao, Zhou, and Zhang]{Gao2012}
Y.~F. Gao, M.~Zhou, and W.~Zhang.
\newblock Novel dispersion properties of one-dimensional photonic crystals
  containing a defect made of twin prisms.
\newblock \emph{Journal of Russian Laser Research}, 33\penalty0 (3):\penalty0
  211--216, May 2012.
\newblock ISSN 1573-8760.
\newblock \doi{10.1007/s10946-012-9274-y}.
\newblock URL \url{https://doi.org/10.1007/s10946-012-9274-y}.

\bibitem[Tsakmakidis et~al.(2007)Tsakmakidis, Boardman, and
  Hess]{Tsakmakidis2007}
K.~L. Tsakmakidis, A.~D. Boardman, and O.~Hess.
\newblock `trapped rainbow' storage of light in metamaterials.
\newblock \emph{Nature}, 450\penalty0 (7168):\penalty0 397--401, Nov 2007.
\newblock ISSN 1476-4687.
\newblock \doi{10.1038/nature06285}.
\newblock URL \url{https://doi.org/10.1038/nature06285}.

\bibitem[Gan et~al.(2009)Gan, Ding, and Bartoli]{Gan2009}
Q.~Gan, Y.~J. Ding, and F.~J. Bartoli.
\newblock ``rainbow'' trapping and releasing at telecommunication wavelengths.
\newblock \emph{Phys. Rev. Lett.}, 102:\penalty0 056801, Feb 2009.
\newblock \doi{10.1103/PhysRevLett.102.056801}.
\newblock URL \url{https://link.aps.org/doi/10.1103/PhysRevLett.102.056801}.

\bibitem[Liu et~al.(2017)Liu, Wang, Han, Shao, Fang, Zhang, Huang, Zhang, and
  Hao]{Liu2017}
Y.~Liu, Y.~Wang, G.~Han, Y.~Shao, C.~Fang, S.~Zhang, Y.~Huang, J.~Zhang, and
  Y.~Hao.
\newblock Engineering rainbow trapping and releasing in ultrathin thz plasmonic
  graded metallic grating strip with thermo-optic material.
\newblock \emph{Opt. Express}, 25\penalty0 (2):\penalty0 1278--1287, Jan 2017.
\newblock \doi{10.1364/OE.25.001278}.
\newblock URL \url{https://opg.optica.org/oe/abstract.cfm?URI=oe-25-2-1278}.

\bibitem[Liu et~al.(2018)Liu, Kanyang, Han, Shao, Fang, Huang, Zhang, Zhang,
  and Hao]{Liu2018}
Y.~Liu, R.~Kanyang, G.~Han, Y.~Shao, C.~Fang, Y.~Huang, S.~Zhang, J.~Zhang, and
  Y.~Hao.
\newblock Rainbow trapping in highly doped silicon graded grating strip at the
  terahertz range.
\newblock \emph{IEEE Photonics Journal}, 10\penalty0 (3):\penalty0 1--9, 2018.
\newblock \doi{10.1109/JPHOT.2018.2816566}.

\bibitem[Arreola-Lucas et~al.(2019)Arreola-Lucas, Báez, Cervera, Climente,
  Méndez-Sánchez, and Sánchez-Dehesa]{ArreolaLucas2019}
A.~Arreola-Lucas, G.~Báez, F.~Cervera, A.~Climente, R.~A. Méndez-Sánchez,
  and J.~Sánchez-Dehesa.
\newblock Experimental evidence of rainbow trapping and bloch oscillations of
  torsional waves in chirped metallic beams.
\newblock \emph{Scientific Reports}, 9:\penalty0 1860, 2019.
\newblock \doi{10.1038/s41598-018-37842-7}.
\newblock URL \url{https://doi.org/10.1038/s41598-018-37842-7}.

\bibitem[Ne\c{s}eli et~al.(2020)Ne\c{s}eli, Bor, Kurt, and Turduev]{Neseli2020}
B.~Ne\c{s}eli, E.~Bor, H.~Kurt, and M.~Turduev.
\newblock Rainbow trapping in a tapered photonic crystal waveguide and its
  application in wavelength demultiplexing effect.
\newblock \emph{J. Opt. Soc. Am. B}, 37\penalty0 (5):\penalty0 1249--1256,
  2020.
\newblock \doi{10.1364/JOSAB.388374}.
\newblock URL
  \url{https://opg.optica.org/josab/abstract.cfm?URI=josab-37-5-1249}.

\bibitem[Ghaderian and Habibzadeh-Sharif(2021)]{Ghaderian2021}
P.~Ghaderian and A.~Habibzadeh-Sharif.
\newblock Rainbow trapping and releasing in graded grating graphene plasmonic
  waveguides.
\newblock \emph{Opt. Express}, 29\penalty0 (3):\penalty0 3996--4009, 2021.
\newblock \doi{10.1364/OE.414982}.
\newblock URL \url{https://opg.optica.org/oe/abstract.cfm?URI=oe-29-3-3996}.

\bibitem[Hayran et~al.(2016)Hayran, Turduev, Botey, Herrero, Staliunas, and
  Kurt]{Hayran:16}
Z.~Hayran, M.~Turduev, M.~Botey, R.~Herrero, K.~Staliunas, and H.~Kurt.
\newblock Numerical and experimental demonstration of a wavelength
  demultiplexer design by point-defect cavity coupled to a tapered photonic
  crystal waveguide.
\newblock \emph{Opt. Lett.}, 41\penalty0 (1):\penalty0 119--122, Jan 2016.
\newblock \doi{10.1364/OL.41.000119}.
\newblock URL \url{https://opg.optica.org/ol/abstract.cfm?URI=ol-41-1-119}.

\bibitem[Johnson and Joannopoulos(2001)]{mpb}
S.~G. Johnson and J.~D. Joannopoulos.
\newblock Block-iterative frequency-domain methods for maxwell's equations in a
  planewave basis.
\newblock \emph{Opt. Express}, 8\penalty0 (3):\penalty0 173--190, 2001.
\newblock URL \url{http://www.opticsexpress.org/abstract.cfm?URI=OPEX-8-3-173}.

\bibitem[Taflove and Hagness(2005)]{taf05}
A.~Taflove and S.~C. Hagness.
\newblock \emph{Computational Electrodynamics: The Finite-Difference
  Time-Domain Method}.
\newblock Artech House, Norwood, 3rd edition, 2005.

\bibitem[Oskooi et~al.(2010)Oskooi, Roundy, Ibanescu, Bermel, Joannopoulos, and
  Johnson]{meep}
A.~F. Oskooi, D.~Roundy, M.~Ibanescu, P.~Bermel, J.~D. Joannopoulos, and S.~G.
  Johnson.
\newblock Meep: A flexible free-software package for electromagnetic
  simulations by the fdtd method.
\newblock \emph{Comput. Phys. Commun.}, 181\penalty0 (3):\penalty0 687--702,
  2010.
\newblock ISSN 0010-4655.
\newblock \doi{10.1016/j.cpc.2009.11.008}.
\newblock URL
  \url{https://www.sciencedirect.com/science/article/pii/S001046550900383X}.

\bibitem[{Liu}(2005)]{Liu2005}
J.~M. {Liu}.
\newblock \emph{{Photonic Devices}}.
\newblock Cambridge University Press, 2005.
\newblock ISBN 9780511614255.
\newblock URL \url{https://dl.acm.org/doi/abs/10.5555/1717189}.

\bibitem[Poon et~al.(2004)Poon, Scheuer, Xu, and Yariv]{Poon2004}
J.~K.~S. Poon, J.~Scheuer, Y.~Xu, and A.~Yariv.
\newblock Designing coupled-resonator optical waveguide delay lines.
\newblock \emph{J. Opt. Soc. Am. B}, 21\penalty0 (9):\penalty0 1665--1673, Sep
  2004.
\newblock \doi{10.1364/JOSAB.21.001665}.

\bibitem[Baba et~al.(2009)Baba, Adachi, Ishikura, Hamachi, Sasaki, Kawasaki,
  and Mori]{Baba2009}
T.~Baba, J.~Adachi, N.~Ishikura, Y.~Hamachi, H.~Sasaki, T.~Kawasaki, and
  D.~Mori.
\newblock Dispersion-controlled slow light in photonic crystal waveguides.
\newblock \emph{Proceedings of the Japan Academy. Series B, Physical and
  biological sciences}, 85\penalty0 (10):\penalty0 443--453, 2009.
\newblock \doi{10.2183/pjab.85.443}.

\bibitem[Yuksel et~al.(2023)Yuksel, Oguz, Karakilinc, Berberoglu, Turduev,
  Adak, and Kart]{yuksel2023enhanced}
Z.~M. Yuksel, H.~Oguz, O.~O. Karakilinc, H.~Berberoglu, M.~Turduev, M.~Adak,
  and S.~Ozdemir Kart.
\newblock Enhanced self-collimation effect by low rotational symmetry in
  hexagonal lattice photonic crystals.
\newblock \emph{ArXiv e-prints}, 2023.
\newblock \doi{10.48550/arXiv.2311.11270}.
\newblock URL \url{https://arxiv.org/abs/2311.11270}.

\bibitem[Engelen et~al.(2006)Engelen, Sugimoto, Watanabe, Korterik, Ikeda, van
  Hulst, Asakawa, and Kuipers]{Engelen2006}
R.~J.~P. Engelen, Y.~Sugimoto, Y.~Watanabe, J.~P. Korterik, N.~Ikeda, N.~F. van
  Hulst, K.~Asakawa, and L.~Kuipers.
\newblock The effect of higher-order dispersion on slow light propagation in
  photonic crystal waveguides.
\newblock \emph{Optics Express}, 14\penalty0 (4):\penalty0 1658--1672, 2006.
\newblock \doi{10.1364/OE.14.001658}.

\bibitem[Hu et~al.(2013)Hu, Ji, Zeng, Liu, and Gan]{Hu2013}
H.~Hu, D.~Ji, X.~Zeng, K.~Liu, and Q.~Gan.
\newblock Rainbow trapping in hyperbolic metamaterial waveguide.
\newblock \emph{Scientific Reports}, 3\penalty0 (1):\penalty0 1249, Feb 2013.
\newblock \doi{10.1038/srep01249}.
\newblock URL \url{https://doi.org/10.1038/srep01249}.

\end{thebibliography}

\end{document}